# Cascade Freezing of Supercooled Water Droplet Collectives

*Gustav Graeber[1], Valentin Dolder[1], Thomas M. Schutzius[1,*], Dimos Poulikakos[1,*]*

[1] Laboratory of Thermodynamics in Emerging Technologies, Department of Mechanical and Process Engineering, ETH Zurich, Sonneggstrasse 3, CH-8092 Zurich, Switzerland.

* To whom correspondence should be addressed.

Short title: Cascade Freezing of Water Droplet Collectives

Prof. Dimos Poulikakos
ETH Zurich
Laboratory of Thermodynamics in Emerging Technologies
Sonneggstrasse 3, ML J 36
CH-8092 Zürich
SWITZERLAND
Phone: +41 44 632 27 38
Fax: +41 44 632 11 76
dpoulikakos@ethz.ch

Dr. Thomas M. Schutzius
ETH Zurich
Laboratory of Thermodynamics in Emerging Technologies
Sonneggstrasse 3, ML J 38
CH-8092 Zürich
SWITZERLAND
Phone: +41 44 632 46 04
thomschu@ethz.ch

**Abstract**

Surface icing affects the safety and performance of numerous processes in technology. Previous studies mostly investigated freezing of individual droplets. The interaction among multiple droplets during freezing is investigated less, especially on nanotextured icephobic surfaces, despite its practical importance as water droplets never appear in isolation, but in groups. Here we show that freezing of a supercooled droplet leads to spontaneous self-heating and induces strong vaporization. The resulting, rapidly propagating vapor front causes immediate cascading freezing of neighboring supercooled droplets upon reaching them. We put forth the explanation that, as the vapor approaches cold neighboring droplets, it can lead to local supersaturation and formation of airborne microscopic ice crystals, which act as freezing nucleation sites. The sequential triggering and propagation of this mechanism results in the rapid freezing of an entire droplet ensemble resulting in ice coverage of the nanotextured surface. Although cascade freezing is observed in a low-pressure environment, it introduces an unexpected pathway of freezing propagation that can be crucial for the performance of rationally designed icephobic surfaces.

Freezing of water droplets and surface icing occur with great frequency in nature and are relevant to the safety and performance of numerous processes in technology.[1,2] This has served as a driver to research into icephobic surfaces, which can passively prevent ice accumulation by different pathways. Through rational surface micro/nanotexturing of hydrophobic materials, one can fabricate superhydrophobic surfaces[3–7] that can shed supercooled water before it freezes due to high droplet mobility, delay freezing[8,9] or reduce ice adhesion.[10–13] Previous work also investigated fundamental freezing phenomena highlighting the critical role of environmental conditions on the freezing outcome.[14–16] While we now have a much improved understanding of the freezing physics for single droplets,[7–16] relatively few studies have investigated the freezing of droplet collectives on micro/nanoengineered icephobic surfaces and the possible emergence of a group dynamics in freezing. This is a very important aspect of the entire icing phenomenon, since droplets do not appear in isolation. Previously, it was demonstrated that freezing can propagate on surfaces by growing frost-halos,[17] ice-bridging,[18–21] shattering of exploding droplets,[22] and ice shrapnels.[23] Here we report and investigate an unexplored mechanism of cascade freezing amongst supercooled droplets. Due to their great potential to serve as icephobic materials, it is plausible to perform our study on micro/nanotextured icephobic surfaces. We find that cascade freezing is caused by airborne vapor boluses, generated and rapidly propagated within a fraction of a second into the surroundings by the spontaneous recalescent freezing of supercooled droplets. We propose that when this vapor bolus reaches the neighboring supercooled droplets it can cause local supersaturation and spontaneous formation of microscopic ice crystals by heterogeneous nucleation condensation on airborne dust and subsequent solidification, triggering nucleation by contact freezing of the neighboring droplet[24] resulting in a freezing cascade. To understand the different possibilities of freezing propagation on surfaces is critical in order to design icephobic materials.

**Results and Discussion**

We investigate the freezing of millimeter-sized water droplets and water droplet arrays resting on solid surfaces under dry ($RH_\infty \leq 3\%$), low-pressure environmental conditions ($p_\infty \approx 3\,\text{mbar}$) at room temperature ($T_\infty \approx 24\,°\text{C}$) (see Methods section). For most of the work we use a spray-coated, superhydrophobic surface with a micro/nanotexture as such spray-coated engineered surfaces have been demonstrated to be promising candidates for the development of icephobic materials[15] (see Methods section and Figure S1 for micrographs of the surface). In the dry, low-pressure environment the sessile droplets quickly self-supercool by evaporation down to a temperature of $T_F \approx -15\,°\text{C}$ where they spontaneously freeze (see Figure S2). Supercooled droplet freezing is a two-step process: In the first stage (recalescence), which only lasts $\approx 0.01\,\text{s}$, the droplet partially solidifies to a slushy mixture of liquid water and ice, simultaneously heating up to the equilibrium freezing temperature of $0\,°\text{C}$ due to release of latent heat.[2] Initially the supercooled droplet is visually transparent, and after the first stage of freezing, it turns opaque. In the second freezing stage, which is usually much slower (lasting seconds), further heat removal leads to complete solidification of the droplet. The sudden heating of the supercooled droplet during recalescence explained above, results in a sharp increase of the droplet evaporation rate,[17] which has been shown to be strong enough to even trigger self-levitation of freezing droplets on rationally designed surface textures[15] or cause frost-halo formation.[17]

Figure 1 shows an image sequence of ten supercooled water droplets freezing on a superhydrophobic surface. Interestingly, within 0.15 s all ten droplets nucleate. High-speed photography reveals that one droplet freezes first (here it is the one on the top right) and starts a freezing cascade in the neighboring supercooled droplets. Subsequent to nucleation, the droplets self-levitate. This levitation mechanism of individual droplets has been reported elsewhere and is not the focus of the present study.[15] Figure S3 and Video S2 show a freezing cascade from a side-view perspective on the spray-coated superhydrophobic surface as used in Figure 1. Figure S4 shows a side-view image sequence of cascade freezing on a single-tier nanotextured glass. The similarity in the freezing behavior of the droplets on the two surfaces is evident.

Figure 2(a) shows the sequential freezing of two droplets of initial droplet volume, $V_D$, at an inter-droplet distance, $s_D$. We recorded the droplet freezing times of the first droplet to freeze, $t_{F1}$, and the second droplet to freeze, $t_{F2}$, which are defined as the time elapsing between the chamber pressure falling below 112.5 mbar (which is the upper limit of the pressure sensor scale) and the start of nucleation of the respective droplet. Keeping $s_D$ constant at $5.0\pm0.3$ mm, we performed this experiment 10 times for three different values of $V_D$, 5 µL, 10 µL and 15 µL, finding an average freezing delay time, $\Delta t_F = t_{F2} - t_{F1}$, of $0.06\pm0.05$ s, $0.05\pm0.05$ s and $0.03\pm0.01$ s, respectively (for detailed data see Figure S5). Remarkably, 26 out of 30 droplet pairs froze with a freezing delay time of $\Delta t_F \leq 0.1$ s. We found that the values of $t_{F1}$ for a given $V_D$ have a normal distribution (Anderson-Darling test, significance level 0.05, see Methods section). Figure 2(b) shows the probability densities of the normal distributions fitted to the experimentally measured values of $t_{F1}$ for the different $V_D$. Based on these probability densities of $t_{F1}$, we compute the probability of two droplets independently freezing (without any interaction between the droplets), $\Psi$, within equal or less than a certain $\Delta t_F$ (see Methods section). In Figure 2(c) we plot $\Psi$ vs $\Delta t_F$ for the same three droplet volumes as in Figure 2(b). Based on this analysis we found that $\Psi \leq 0.03$ for $\Delta t_F \leq 0.1$ s for all $V_D$, meaning that the probability of two droplets freezing independently of each other within 0.1 s is 3%. Figure 2(d) shows a plot of occurrence vs $s_D$ for three different $V_D$, with occurrence being defined by the observed behavior of the droplets: cascade freezing, ice shattering, and no freezing. The time for observing the occurrence is limited to $\Delta t_\text{F} \leq 0.1\,\text{s}$. We see that for a range of $s_\text{D}$ and $V_\text{D}$ the second droplet freezes with high-probability in spite of the small value of $\Delta t_\text{F}$, indicating that the freezing droplet pair cannot be attributed to chance. It is also important to note that for $s_\text{D} < 30\,\text{mm}$ ice shattering is not the dominant mechanism responsible for causing the neighboring droplet to freeze. We find that the smaller $s_\text{D}$ and the larger $V_\text{D}$, the more we observe cascade freezing and that for sufficiently large $s_\text{D}$, we do not observe cascade freezing anymore. We tested this specific range of $V_\text{D}$ as these droplet sizes are common in technical applications and nature.[25]

To further elucidate the freezing cascade mechanism, we placed a solid metal barrier between the two droplets as shown in Figure 2(e) and followed the previous protocol. Figure 2(f) shows an image sequence of two droplets freezing with a solid barrier in between them, prohibiting the freezing of the droplet that freezes first from triggering nucleation of its neighboring droplet. While the average $t_\text{F1}$ for the different $V_\text{D}$ does not change significantly by adding the barrier, the average $\Delta t_\text{F}$ increases by more than two orders of magnitude to $\Delta t_\text{F} \sim 10\,\text{s}$ (see Figure S5 for details). We did not observe freezing propagation anymore, when the solid barrier was placed between the droplets.

We want to discuss the effects that might be responsible for the freezing of a droplet to cause a neighboring droplet to freeze, termed here freezing propagation, in order to explain the "cascade freezing" mechanism that we observed. For our study, we can exclude freezing propagation mechanisms based on frost halos[17] and ice-bridging[18–21] as we do not observe frost halo growth or ice bridges, and the freezing delay observed in our experiments is multiple orders of magnitude shorter than freezing propagation based on the other effects. We can further exclude the shattering of entire, exploding droplets,[22] as the time in our experiments for a freezing droplet to explode is an order of magnitude longer than the limit of cascade freezing of 0.1 s (see Video S4 which shows freezing propagation based on droplet explosion, where $\Delta t_F > 1\,\text{s}$). It has been shown that the shattering of ice spicules, which form during the second stage of freezing, can result in the ejection of rapidly propagating ice particles of a size of about 100 μm, which can cause neighboring droplets to freeze.[22] We carefully analyzed all experiments from Figure 2 where the two droplets froze with $\Delta t_F \leq 0.1\,\text{s}$. Using our high-zoom observations, which allowed identifying objects with a size of approximately 10 μm, we saw ice spicule formation and ejection of ice splinters before nucleation of both droplets in only 5 % of the freezing propagation cases. Even larger magnification videos, did not show ice splinters to play a significant role (see Video S5). Based on these observations we exclude exploding ice spicules as a main contributor to the freezing propagation in our experiments. Here we postulate that it is the airborne vapor released during recalescence by the first droplet freezing, that is mainly responsible for the freezing cascade mechanism we observed.

We further characterized the released vapor bolus during recalescence freezing and present the results in Figure 3. We found that the released vapor is strong enough to visibly deform a thin steel cantilever beam positioned above the freezing droplet (see Figure 3(a)-(d) and Video S6). In line with our previous discussion, no ejected solid ice particles were observed to be responsible for deforming the beam. Using two thermocouples in the proximity of the droplet as shown in Figure 3(e), we measured the vapor bolus speed to be $u_V \approx 0.5\,\text{m/s}$ (see Figure 3(g), (f)). Based on the environmental conditions ($p_\infty \approx 3\,\text{mbar}$, $T_\infty \approx 24\,°\text{C}$) we estimate the diffusivity of water vapor in air to be $D_V = 0.211 \cdot (T_\infty / T_{\text{ref}})^{1.94} \cdot (p_{\text{ref}} / p_\infty) = 84\,\text{cm}^2/\text{s}$ (Ref. 25, with $T_{\text{ref}} = 273.15\,\text{K}$ and $p_{\text{ref}} = 1013.25\,\text{mbar}$). The corresponding diffusion speed is then $u_D = x/t = 2 \cdot D_V / x \approx 1\,\text{m/s}$, where $x = 15\,\text{mm}$, which corresponds to the average distance between the thermocouples in the experiments (see Figure 3(g)). Comparing the orders of magnitude of $u_V$ and $u_D$, we find that, at the radial distance from the droplet where we measure the speed ($\approx 10$ mm from the droplet surface), the vapor transport mechanism can be assumed to be diffusion. The low-pressure environment leads to high diffusive vapor speeds and allows the released vapor to quickly spread. In contrast, for a system at $24\,°C$ and 1013 mbar, the corresponding diffusion speed is expected to be 0.003 m/s, which is markedly slower, reducing inter-droplet interactions in a recalescence freezing event. We further attempted to measure the temperature of the released vapor, $T_V$, using two thermocouples of different size positioned at a distance of approximately 1 mm from the freezing droplet (see Fig. 3(h)). After averaging over 11 independent experiments and correcting for the thermal inertia of the thermocouples, we find that both thermocouples measure a similar temperature profile in time, as shown in Fig. 3(i). We use the minimum of this temperature profile as an estimate for the vapor temperature at the location of the thermocouples and find $T_V \approx 8\,°C$ (see Methods section). Due to uncertainties in the estimation of the effects of thermal inertia and heat conduction along the thermocouple wires, this value is only a rough estimate. It is expected that the vapor at its release location is at the equilibrium freezing temperature of $0°C$. The chamber temperature away from the droplets is close to room temperature. Hence this measurement appears to be plausible.

Figure 4 shows a high-speed infrared recording of the cascade freezing effect between two water droplets from a top-view perspective performed on a poly(methyl methacrylate) (PMMA) substrate. PMMA allows us to observe the entire droplet free surface and triple line from the top-view perspective. In the first frame, at $t = 0$ s, both droplets are supercooled to a temperature $T_F \approx -15\,°C$. The droplet on the left spontaneously nucleates, which results in a local warming due to the release of latent heat. The pronounced contrast of the infrared recording allows us to precisely locate the freezing front. After about 0.02 s the droplet on the left completes its first stage of freezing. Simultaneously, the neighboring droplet nucleates. The very short freezing delay which is insufficient for internal pressure built-up and fracturing, and the visible absence of ice spicules further underpin, that the ejection of splinters is not responsible for the freezing propagation. The nucleation from the droplet free surface supports that the propagation of freezing is based on an airborne input.

From Figure 4, the temperature of the low-pressure gas surrounding the supercooled droplets cannot be assessed. We used a fine thermocouple to estimate the temperature of the gas next to a supercooled droplet shortly before spontaneous nucleation of the droplet. We found that the gas close to the supercooled droplet, within a radial distance of $r = r_\infty \approx 5$ mm measured from the droplet center is colder than $T_\infty$ (see Figure S6). The relatively low heat exchange between the thermocouple and the low-pressure gas combined with heat conduction along the thermocouple wires hinder an accurate measurement of the temperature field around the supercooled droplet. From the measurement we can learn though, that there is a cold region around the supercooled droplet, which extends in radial direction at least towards the radius $r = r_\infty$ (see Figure S6). With this information we can perform a simplified heat transfer analysis, assuming steady state, one-dimensional, radial heat conduction in the gas. We neglect effects of natural convection, justified by a low Rayleigh number

$Ra = \frac{g \cdot \beta}{\nu \cdot \alpha}(T_\infty - T_F) \cdot x^3 \approx 0.1$ (Ref. 26) computed with the gravitational constant $g = 9.81 \,\mathrm{m \cdot s^{-2}}$, assuming $\beta = (0.5 \cdot (T_\infty + T_F))^{-1}$, using $x = r_\infty = 5$ mm as a characteristic length, and kinematic viscosity $\nu = 0.001315 \,\mathrm{m^2 s^{-1}}$ and thermal diffusivity $\alpha = 0.001308 \,\mathrm{m^2 s^{-1}}$ of saturated water vapor evaluated at the average of $T_F$ and $T_\infty$ (Ref. 27). From the simplified conduction analysis in saturated vapor with a thermal conductivity of $k(T = 0.01\,°\mathrm{C}) = 0.0168 \,\mathrm{W/(mK)}$ (Ref. 27), assuming $T(r = r_\infty) = T_\infty \approx 24\,°\mathrm{C}$ at a radial position $r = r_\infty \approx 5\,\text{mm}$ and $T(r = r_S) = T_F \approx -15\,°\text{C}$ at the droplet free surface at a radial position $r = r_S \approx 1.3\,\text{mm}$, we estimate that the gas above the droplet surface up to a distance of $0.5\,\text{mm}$ is at a temperature below 0°C (see Figure S6).

Based on our observations, measurements, and calculations we have strong evidence supporting the following sequence of events: When a supercooled water droplet freezes, it emits suddenly a vapor bolus to its surrounding environment. After the emitted vapor bolus departs the surface of the first droplet, it diffuses towards the neighboring droplet, which acts as a heat sink to the vapor. Assuming plausibly that the incoming water vapor is saturated with respect to liquid water at a temperature close to 0 °C and that the neighboring supercooled droplet cools it down to a temperature close to $-15\,°\text{C}$, we find that locally the saturation level with respect to liquid water reaches $S = p_{V,sat,liq}(T = 0\,°\text{C}) / p_{V,sat,liq}(T = -15\,°\text{C}) = 3.19$ (Ref. 28). These conditions do not support homogenous nucleation of droplets; however, as the experiments are performed in a standard laboratory environment, it is reasonable to assume that there are airborne dust particles that can strongly reduce the free energy barrier for condensation nucleation and facilitate the formation of small droplets on the dust by heterogeneous nucleation. Assuming standard room air,[29] we know that there is a large number of airborne particles of a size of up to $5\,\mu\text{m}$ present and we can estimate a total dust particle surface area per cubic meter of $\rho_{Dust} = 2.7 \cdot 10^{-4}\,\text{m}^2/\text{m}^3$ by summing up all particles multiplied by their respective surface area. The subzero region around the neighboring droplet (with $V_D = 10\,\mu\text{L}$) is $V_{subzero} = 4/3 \cdot \pi \cdot \left[ (r_S + 0.5\,\text{mm})^3 - r_S^3 \right] = 1.6 \cdot 10^{-8}\,\text{m}^3$, so that we estimate the effective dust particle surface area possibly involved in nucleation to be $A_{Dust} = \rho_{Dust} \cdot V_{subzero} = 10^{-12}\,\text{m}^2$. By accounting for the pressure reduction in the chamber, one can conservatively assume that as the pressure is reduced from atmospheric pressure down to about 1 mbar, the number of particles in the chamber will be reduced proportionally to $A_{\text{Dust,min}} = A_{\text{Dust}}/1000 = 10^{-15} \text{ m}^2$.

The rate of nucleation can be computed as $J = K \cdot \exp\left(-\dfrac{\Delta G_C}{k_B \cdot T}\right)$ (Ref. 30), where $k_B$ is the Boltzmann constant, and we assume the relevant temperature to be $T = -15\,°\text{C}$. For condensation the pre-factor is $K = J_0 = 10^{29}\,\dfrac{\text{embryo}}{\text{m}^2 \cdot \text{s}}$ (Ref. 31). The Gibbs free energy barrier in the case of condensation is computed as $\Delta G_C = \dfrac{16\pi \cdot \gamma_{LV}^3 \cdot v_m^2 \cdot f(\theta_{LD})}{3 \cdot (R \cdot T \cdot \ln(S))^2}$, where the interfacial energy between liquid water and vapor is computed as

$\gamma_{LV} = (75.7 - 0.1775\,(T/K)) \cdot 10^{-3}\,\text{N/m}$ (Ref. 31). The molar volume of water is $v_m = 1.802 \cdot 10^{-5}\,\text{m}^3/\text{mol}$ (at $T = 273.16\,\text{K}$ and $p = 1\,\text{bar}$, Ref. 27). $R = 8.314\,\text{J/(molK)}$ is the ideal gas constant. The factor $f(\theta_{LD}) = 0.25 \cdot [2 + \cos(\theta_{LD})] \cdot [1 - \cos(\theta_{LD})]^2$ (Ref. 32) takes into account the contact angle $\theta_{LD}$ between the condensing liquid water and the solid dust particle surface. The effect of the dust particle curvature is not considered here, as the dust particles are much larger than the critical embryo size.[30,33] Mineral dust – including calcite, quartz and clays – is a major contributor to the airborne dust particles.[34] We can thus assume that the water contact angle with the dust is fairly small, $\theta_{LD} \approx 40°$ (Ref. 35). The resulting number of liquid condensation embryos being generated within a time of $\Delta t = 0.1\,\text{s}$ (the time of a freezing cascade) is $N_{\text{Embryo}} = J \cdot A_{\text{Dust,min}} \cdot \Delta t = 10^{11}$. Near the supercooled droplet, at a temperature of $-15\,°\text{C}$, it is plausible that at least one of the so-formed condensation droplets will heterogeneously freeze on the dust surface and form an airborne microscopic ice crystal.

When this formed ice crystal comes in contact with the neighboring supercooled droplet, it will instantly trigger recalescent freezing. Due to optical limitations and the speed of the involved phenomena, we cannot show the microscopic condensation droplets or ice crystals which form next to the supercooled droplet, but we can perform a comparable experiment showing heterogeneous nucleation of airborne droplets in our experimental chamber (see Figure S7 and Video S8). In this experiment we rapidly pump down the environmental chamber filled with standard room air from the initial condition ($T_\infty = 24\,°C$; $p_\infty = 952\,\text{mbar}$; $RH_\infty = 51\,\%$). Approximately 0.5 seconds after opening the valve between vacuum pump and environmental chamber, we observe the formation of microscopic airborne condensation droplets (see Fig. S7). Doing the same experiment when starting from $RH_\infty = 3\,\%$ does not result in any particle formation. At the moment of airborne particles becoming visible, the pressure in the chamber has reached $p_\infty \approx 700\,\text{mbar}$. Assuming an adiabatic and reversible expansion of the gas inside the chamber (with a heat capacity ratio of $\gamma_{\text{air}}(T = 300\,\text{K}) = 1.4$ (Ref. 27)), we estimate that the gas in the chamber has cooled to approximately $T_{\text{Gas,local}} \approx 0\,°C$ at this point, which corresponds to a local saturation of $S_{\text{local}} \approx 2$. We can conclude conservatively that there are airborne particles in the chamber which activate condensation at saturation levels of 2 which is well below the supersaturation we expect during cascade freezing.

The proposed freezing propagation explanation fits well with our experimental findings and observations. It explains why the nucleation of the neighboring droplet starts from the free surface and not from the triple line and why a rigid, impermeable barrier can stop the cascade freezing effect. It is also in line with our finding, that the cascade freezing probability decreases with increasing $s_D$ and decreasing $V_D$. As we found that the vapor propagation can be assumed to be diffusion-driven, increasing $s_D$ will reduce the achievable supersaturation near the neighboring droplet. Based on Fick's law we estimate the extent of the region from the freezing droplet getting affected by the released vapor to be $\Delta x = \dfrac{D_V \cdot p_V}{R_{H_2O} \cdot T_{ref} \cdot j} \approx 4\,\text{mm}$, where we used $D_V = 84\,\text{cm}^2/\text{s}$, $p_V(T=0°C) = 6.1\,\text{mbar}$, $R_{H_2O} = 461.5\,\text{J/(kgK)}$, $T_{ref} = 273.15\,\text{K}$ and $j = 1\,\text{mg/(cm}^2 \cdot \text{s)}$ (Ref. 15). In contrast, increasing $V_D$ (also beyond the range tested in this study) is expected to result in the release of a larger vapor quantity, which leads to the supersaturation of a larger region and facilitates cascade freezing.

To rule out effects of local cooling by dry gas flow[16] or the effects of the dust in the surrounding gas triggering droplet freezing when displaced by the released vapor bolus towards the neighboring droplet,[24] we performed experiments with an electric fan in the environmental chamber directed at the droplets (see Figure S8). The activation of the fan, which results in a gas flow speed of approximately 3 m/s and to a displacement of airborne dust particles onto the droplet free surface, did not markedly change the time a droplet takes to spontaneously nucleate (see Figure S8). Therefore, the dust particles alone without the vapor bolus do not lead to freezing propagation, leaving the emitted vapor bolus and its heterogeneous nucleation as the most likely reason for the observed freezing group dynamics.

**Summary and Conclusions**

In summary, we described the mechanism of cascade freezing amongst evaporatively supercooled neighboring droplets. We proposed that the freezing group dynamics is caused by airborne vapor boluses, which are generated and rapidly propagated into the surroundings during spontaneous recalescent freezing of the supercooled droplets. When this vapor bolus reaches the neighboring supercooled droplets it can cause local supersaturation and spontaneous formation of microscopic ice crystals by heterogeneous nucleation condensation on airborne dust and subsequent solidification, triggering nucleation by contact freezing of the neighboring droplet. The repetition of this series of events results in the observed freezing cascade. The proposed physics clearly distinguishes cascade freezing from other freezing propagation mechanisms.

**Methods**

**Materials**

For the experiments we used deionized water (DIW, Merck Milli-Q direct, resistivity > 18.2 MΩ cm). We obtained 1-Methyl-2-pyrrolidinone (NMP), Poly(methyl methacrylate) (PMMA) powder and poly(vinylidene fluoride) (PVDF) pellets from Sigma Aldrich. We obtained acetone from Thommen-Furler AG and hydrophobic fumed silica nanoparticles (HFS, Aerosil R8200) from Evonik. We used sheets of PMMA in thickness of 175 μm and 1 mm obtained from Schlösser GmbH and transparent float glass microscope slides in thickness of 1 mm from VWR (cut edge, ECN 631-1550). We purchased bare wire fine gage T-type thermocouples from Omega: COCO-001 with wire diameter of 0.025 mm and COCO-003 with wire diameter of 0.075 mm.

**Substrate preparation**

The glass slides were spray coated to create a micro/nanotextured, superhydrophobic surface, which is used for most experiments in this study (see inset in Figure 1 and Figure S1). We used a siphon feed airbrush (Paasche, VL 0316, 0.73 mm head, back-pressure ~3 bar), creating a coating similar to Ref. 36. For the coating we prepared separate solutions of 10 wt.% PMMA in acetone and 10 wt.% PVDF in NMP by dissolving the polymers under slow mechanical mixing for 12 hours at room temperature and at 50 °C, respectively. The coating consists of 1.16 g of HFS, which was mixed with 16.88 g acetone by probe sonication three times for 30 s. To the HFS-acetone suspension we added 1 g of each the PVDF in NMP mixture and the PMMA in acetone mixture and mixed it by shaking. After spray coating, the samples were heated for 10 minutes on a hotplate at 100 °C to evaporate all residual solvents. We measured the advancing and receding contact angles on the as-fabricated spray-coated sample to be $\theta^*_{adv} = 163° \pm 4°$ and $\theta^*_{rec} = 155° \pm 6°$, respectively (averaged over five individual measurements, giving the standard deviation as error). After performing numerous freezing experiments on the sample it did not show any measurable wetting deterioration. We measured $\theta^*_{adv} = 165° \pm 2°$ and $\theta^*_{rec} = 162° \pm 2°$, which is in the range of the as-fabricated sample. The PMMA substrates were used in their as-purchased state, with $\theta^*_{adv} = 86° \pm 2°$ and $\theta^*_{rec} = 59° \pm 2°$. The nanotextured glass surface, which is used only in Figure S4, is fabricated by mask-less reactive ion etching for two hours and subsequent hydrophobization. Details on the surface and its preparation are presented in Ref. 14.

**Statistics**

To obtain quantitative data, experiments were repeated $n$ times and the measurements are given as average value $\pm$ the standard deviation. $n$ is specified in the captions or in the text. We checked if experimental data is drawn from a normal distribution using the Anderson–Darling test at a significance level of 0.05. To compare the population means between two groups where a normal distribution can be assumed, we performed a two-sided two-sampled Student's t test and applied the Welch correction. To compare group populations where we found departures from normality, we used the nonparametric Kruskal-Wallis ANOVA test. We indicated significant differences by asterisks (* $P < 0.05$, ** $P < 0.01$ and *** $P < 0.001$), while non-significant differences (significance level 0.05) are marked by "n.s.".

To find the likelihood $\Psi$ in Fig. 2(c), defined as the probability of the two droplets of a droplet pair in the experiment to independently (by coincidence) freeze within a certain freezing delay time $\Delta t_F$, we performed the following steps: We first ensured that the recorded values for $t_{F1}$ were normally distributed (Anderson-Darling test). We fitted a normal distribution to our experimental data and obtained the probability densities as shown in Fig. 2(b), with a mean freezing time $\overline{t}_{F1}$ and a standard deviation $\sigma_{F1}$. As we investigate here if the two droplets of the droplet pair froze independently, the freezing behavior of both droplets is assumed to behave according to the normal distribution of the recorded $t_{F1}$ values. We evaluated the probability density functions using the normal cumulative distribution function (normcdf) in MATLAB, which we represent here by $F$ and is defined as

$$F(t) = \frac{1}{\sqrt{2\pi}} \cdot \int_{-\infty}^{t} \exp\left[-0.5 \cdot \left(\frac{t - \overline{t}_{F1}}{\sigma_{F1}}\right)\right] dt.$$

$t$ is the moment in time at which the function $F$ is evaluated. We introduce a time vector, $T$, ranging from $t_{low} = \overline{t}_{F1} - 5 \cdot \sigma_{F1}$ up to $t_{up} = \overline{t}_{F1} + 5 \cdot \sigma_{F1}$ in steps of $\delta_t = 0.05$ s (where the effect of the choice of $\delta_t$ is negligible) with $n_T = (t_{up} - t_{low})/\delta_t$ entries. $\Psi$ is then computed as

$$\Psi = \sum_{i=1}^{n_T} \{[F(T_i + \delta_t) - F(T_i)] \cdot [F(T_i + \Delta t_F) - F(T_i)]\}.$$

In this equation the first part $[F(T_i + \delta_t) - F(T_i)]$ represents the specific probability of the first-freezing droplet to spontaneously freeze after a certain $t_{F1}$, while the second part of the equation $[F(T_i + \Delta t_F) - F(T_i)]$ represents the specific probability of the second-freezing droplet to spontaneously freeze after the first-freezing droplet in a time interval of this specific $t_{F1}$ and $t_{F1} + \Delta t_F$. By varying $\Delta t_F$ from 0 s to 0.2 s, we obtain Fig. 2(c), which plots the resulting $\Psi$ as a function of $\Delta t_F$.

**Experimental procedure**

The environmental chamber was purged with gaseous nitrogen to reach a dry condition with $RH_\infty \leq 3\%$. Subsequently, the valve to the nitrogen inlet was closed again and the valve a connected vacuum pump was opened in order to reduce the pressure inside the chamber. A detailed description of the experimental setup can be found elsewhere[14].

To measure the vapor speed, we used two COCO-001 thermocouples. We placed one of them < 2 mm from the droplet surface (thermocouple T1), and the other one at a larger distance on the opposing side of the droplet (thermocouple T2). We use the signals of the thermocouples before recalescence freezing (gray area in Fig. 3(f)) to compute the mean and the standard deviation of the temperature measurement before freezing. We specify a temperature corridor based on the mean ± six times the standard deviation. We define the moment when the thermocouple temperature signal leaves this temperature corridor, as the time when the vapor reached the thermocouple position. The uncertainty in the vapor speed measurements results from the effective thermocouple recording sampling rate of ~10 ms.

For the vapor temperature measurements, we used a thick (wire diameters $d_{T3} = 0.075$ mm) and a thin thermocouple ($d_{T4} = 0.025$ mm), which we term T3 and T4, respectively. Their measurement tips are spherical beads, with a diameter of approximately 2.5 times their wire diameter according to the manufacturer's specifications. At this ratio between bead and wire diameter, the thermocouple time constant $\tau_T$ can be estimated based on the wire of the thermocouple itself, due to the significant heat conduction between the wire and the bead.[37] Neglecting radiation (due to low temperatures) and assuming a uniform cross-sectional temperature distribution in the thin wires (due to small Biot number), the dynamic response of the thermocouples can be modeled as a first order system with $T_V - T_T = \tau_T \frac{\partial T_T}{\partial t}$. Here $T_V$ and $T_T$ are the temperature of the vapor and the measured temperature by the thermocouple, respectively. For our cylindrical wire $\tau_T = \frac{\rho_T \cdot d_T \cdot c_T}{4 \cdot h_T}$, where $\rho_T$, $c_T$, and $h_T$ are the density, the specific heat capacity and the heat transfer coefficient of the thermocouple. The thermocouple is made from copper (Cu) and constantan (Co), so that $\rho_T \approx \rho_{Cu} \approx \rho_{Co} \approx 8900 \text{ kg/m}^3$ (Ref. 27) and $c_T \approx c_{Cu} \approx c_{Co} \approx 390 \text{ J/(kg} \cdot \text{K)}$ (Ref. 27). $h_T$ is found based on $h_T = \text{Nu} \cdot k_V / d_T$, where the thermal conductivity of saturated vapor at 0.01°C is $k_V = 0.016761 \text{ W/(mK)}$ (Ref. 27) and Nu is the average Nusselt number. To estimate Nu, we first compute the Reynolds number $\text{Re}_T = d_T \cdot u_V \cdot \rho_V / \mu_V$ and the Prandtl number $\text{Pr} = c_{p,V} \cdot \mu_V / k_V$. By inserting appropriate values (vapor speed $u_V = 0.5 \text{ m/s}$, density $\rho_V = 0.0048546 \text{ kg} \cdot \text{m}^{-3}$, dynamic viscosity $\mu_V = 8.9458 \cdot 10^{-6} \text{ Pa} \cdot \text{s}$, specific heat capacity $c_{p,V} = 1884.4 \text{ J/(kgK)}$ for saturation and an assumed temperature of 0 °C) (Ref. 27) we obtain $\text{Re}_{T3} = 0.0204$, $\text{Re}_{T4} = 0.0068$ and $\text{Pr} = 1.0058$. For these very low Reynolds numbers, Nu based on $d_T$ can be estimated by $\text{Nu} = 1/(0.8237 - \ln((\text{Re}_T \cdot \text{Pr})^{0.5}))$ (Ref. 38). We obtain $\text{Nu}_{TC3} = 0.3612$, $\text{Nu}_{TC4} = 0.3014$, $\tau_{TC3} = 0.8012 \text{ s}$ and $\tau_{T4} = 0.1067 \text{ s}$. We performed 11 individual experiments placing the two thermocouples at a distance from the droplet surface of $0.8 \pm 0.2 \text{ mm}$ (mean ± standard deviation). The distance between the droplet and T3, and the distance between the droplet and T4 differed by less than 0.1 mm in all experiments. Analyzing the high-speed videos (2000 frames per second; 2000 synchronized data points per channel per second) we found the moment in time when droplet freezing started and set this moment to be $t = 0 \text{ s}$. We averaged all 11 experiments in time, obtaining one average curve for T3 and one average curve for T4. Using MATLAB we computed for all moments in time $i$ a second order polynomial best fit to the neighboring $i - 60 \ldots i + 60$ moments in time, obtaining a smoothened temperature signal $T_T$ and by taking the first derivative also obtaining the derivative in all moments in time $\frac{\partial T_T}{\partial t}$. Using $\tau_T$ we can then compute the response time compensated temperatures by: $T_V = T_T + \tau_T \frac{\partial T_T}{\partial t}$. To estimate the error, we compute the standard deviation from the mean during time-averaging the 11 experiments. We perform the same analysis as before to the average temperature curve minus the standard deviation, and to the average temperature curve plus the standard deviation.

**Acknowledgements**

Partial support of the European Research Council under Advanced Grant 669908 (INTICE) and the Swiss National Science Foundation under Grant 162565 is acknowledged. We thank T. Vasileiou, C. Walker and C. Hail for helpful discussions, S. Luh for support in initial experiments, J. Vidic and P. Feusi for assistance in experimental setup construction and U. Drechsler for advice on surface fabrication.

**Author contributions**

D.P. and T.M.S. designed research; G.G. and V.D. performed research; G.G., T.M.S. and V.D. analyzed data; and G.G., T.M.S. and D.P. wrote the paper.

**Competing financial interests**

The authors declare no competing financial interests.

**Supporting Information**

The Supporting Information is available free of charge on the ACS Publications website. It contains eight Supporting Information Figures: Figure S1 Micrographs obtained by scanning electron microscopy of the spray-coated superhydrophobic surface; Figure S2 Freezing time and temperature at freezing initiation; Figure S3 Side-view image sequence of cascade freezing on the spray-coated surface; Figure S4 Side-view image sequence of cascade freezing on a single-tier, nanotextured superhydrophobic surface; Figure S5 Effect of adding a barrier between two neighboring droplets on freezing time and freezing delay; Figure S6 Temperature field around a supercooled droplet; Figure S7 Nucleation of condensation droplets in the environmental chamber; Figure S8 Effect of a dry gas flow on droplet freezing (PDF).

Video S1. Top-view perspective of cascade freezing of 10 evaporatively supercooled water droplets resting on a superhydrophobic substrate.

Video S2. Cascade freezing of evaporatively supercooled water droplets resting on a superhydrophobic substrate.

Video S3. Recalescent freezing of neighboring droplets: Without and with a solid barrier between the supercooled droplets.

Video S4. Freezing propagation based on droplet explosion, where $\Delta t_F > 1\,\text{s}$.

Video S5. Large magnification video of cascade freezing of two evaporatively supercooled water droplets.

Video S6. Analyzing the released vapor during recalescence freezing.

Video S7. Cascade freezing of two evaporatively supercooled water droplets. Infrared photography.

Video S8. Rapid pump down of the environmental chamber starting from ambient conditions resulting in the formation of microscopic airborne water droplets.

**References**

(1)  Kreder, M. J.; Alvarenga, J.; Kim, P.; Aizenberg, J. Design of Anti-Icing Surfaces: Smooth, Textured or Slippery? *Nat. Rev. Mater.* **2016**, *1*, 15003.

(2)  Schutzius, T. M.; Jung, S.; Maitra, T.; Eberle, P.; Antonini, C.; Stamatopoulos, C.; Poulikakos, D. Physics of Icing and Rational Design of Surfaces with Extraordinary Icephobicity. *Langmuir* **2015**, *31*, 4807–4821.

(3)  Lafuma, A.; Quéré, D. Superhydrophobic States. *Nat. Mater.* **2003**, *2*, 457–460.

(4)  Deng, X.; Mammen, L.; Butt, H. J.; Vollmer, D. Candle Soot as a Template for a Transparent Robust Superamphiphobic Coating. *Science* **2012**, *335*, 67–70.

(5)  Verho, T.; Bower, C.; Andrew, P.; Franssila, S.; Ikkala, O.; Ras, R. H. A. Mechanically Durable Superhydrophobic Surfaces. *Adv. Mater.* **2011**, *23*, 673–678.

(6)  Peng, C.; Chen, Z.; Tiwari, M. K. All-Organic Superhydrophobic Coatings with Mechanochemical Robustness and Liquid Impalement Resistance. *Nat. Mater.* **2018**, *17*, 355–360.

(7)  Wong, T.-S.; Kang, S. H.; Tang, S. K. Y.; Smythe, E. J.; Hatton, B. D.; Grinthal, A.; Aizenberg, J. Bioinspired Self-Repairing Slippery Surfaces with Pressure-Stable Omniphobicity. *Nature* **2011**, *477*, 443–447.

(8)  Tourkine, P.; Le Merrer, M.; Quere, D. Delayed Freezing on Water Repellent Materials. *Langmuir* **2009**, *25*, 7214–7216.

(9)  Eberle, P.; Tiwari, M. K.; Maitra, T.; Poulikakos, D. Rational Nanostructuring of Surfaces for Extraordinary Icephobicity. *Nanoscale* **2014**, *6* (9), 4874.

(10) Davis, A.; Yeong, Y. H.; Steele, A.; Bayer, I. S.; Loth, E. Superhydrophobic

Nanocomposite Surface Topography and Ice Adhesion. *ACS Appl. Mater. Interfaces* **2014**, *6*, 9272–9279.

(11) Varanasi, K. K.; Deng, T.; Smith, J. D.; Hsu, M.; Bhate, N. Frost Formation and Ice Adhesion on Superhydrophobic Surfaces. *Appl. Phys. Lett.* **2010**, *97*, 234102.

(12) Subramanyam, S. B.; Rykaczewski, K.; Varanasi, K. K. Ice Adhesion on Lubricant-Impregnated Textured Surfaces. *Langmuir* **2013**, *29*, 13414–13418.

(13) Meuler, A. J.; Smith, J. D.; Varanasi, K. K.; Mabry, J. M.; McKinley, G. H.; Cohen, R. E. Relationships between Water Wettability and Ice Adhesion. *ACS Appl. Mater. Interfaces* **2010**, *2*, 3100–3110.

(14) Graeber, G.; Schutzius, T. M.; Eghlidi, H.; Poulikakos, D. Spontaneous Self-Dislodging of Freezing Water Droplets and the Role of Wettability. *Proc. Natl. Acad. Sci. U. S. A.* **2017**, *114*, 11040–11045.

(15) Schutzius, T. M.; Jung, S.; Maitra, T.; Graeber, G.; Köhme, M.; Poulikakos, D. Spontaneous Droplet Trampolining on Rigid Superhydrophobic Surfaces. *Nature* **2015**, *527*, 82–85.

(16) Jung, S.; Tiwari, M. K.; Doan, N. V.; Poulikakos, D. Mechanism of Supercooled Droplet Freezing on Surfaces. *Nat. Commun.* **2012**, *3*, 615.

(17) Jung, S.; Tiwari, M. K.; Poulikakos, D. Frost Halos from Supercooled Water Droplets. *Proc. Natl. Acad. Sci. U. S. A.* **2012**, *109*, 16073–16078.

(18) Nath, S.; Ahmadi, S. F.; Boreyko, J. B. A Review of Condensation Frosting. *Nanoscale Microscale Thermophys. Eng.* **2017**, *21*, 81–101.

(19) Zhao, Y.; Wang, R.; Yang, C. Interdroplet Freezing Wave Propagation of

Condensation Frosting on Micropillar Patterned Superhydrophobic Surfaces of Varying Pitches. *Int. J. Heat Mass Transfer* **2017**, *108*, 1048–1056.

(20) Boreyko, J. B.; Collier, C. P. Delayed Frost Growth on Jumping-Drop Superhydrophobic Surfaces. *ACS Nano* **2013**, *7*, 1618–1627.

(21) Hao, Q.; Pang, Y.; Zhao, Y.; Zhang, J.; Feng, J.; Yao, S. Mechanism of Delayed Frost Growth on Superhydrophobic Surfaces with Jumping Condensates: More than Interdrop Freezing. *Langmuir* **2014**, *30*, 15416–15422.

(22) Wildeman, S.; Sterl, S.; Sun, C.; Lohse, D. Fast Dynamics of Water Droplets Freezing from the Outside in. *Phys. Rev. Lett.* **2017**, *118,* 084101.

(23) Boreyko, J. B.; Hansen, R. R.; Murphy, K. R.; Nath, S.; Retterer, S. T.; Collier, C. P. Controlling Condensation and Frost Growth with Chemical Micropatterns. *Sci. Rep.* **2016**, *6,* 19131.

(24) Ladino Moreno, L. A.; Stetzer, O.; Lohmann, U. Contact Freezing: A Review of Experimental Studies. *Atmos. Chem. Phys.* **2013**, *13,* 9745–9769.

(25) Pruppacher, H. R.; Klett, J. D. Microphysics of Clouds and Precipitation. In *Atmospheric and Oceanographic Sciences Library*; Mysak, L. A., Hamilton, K., Eds.; Springer: Heidelberg, 1998.

(26) Incropera, F. P.; DeWitt, D. P.; Bergman, T. L.; Lavine, A. S. *Fundamentals of Heat and Mass Transfer,* 6th ed.; John Wiley & Sons: Hoboken, NJ, 2006.

(27) *CRC Handbook of Chemistry and Physics*, 97th ed.; Haynes, W. M., Ed.; CRC Press/Taylor & Francis: Boca Raton, FL, 2016.

(28) Murphy, D. M.; Koop, T. Review of the Vapour Pressures of Ice and Supercooled

Water for Atmospheric Applications. *Q. J. R. Meteorol. Soc.* **2005**, *131*, 1539–1565.

(29) ISO 14644-1:2015 Classification of Air Cleanliness by Particle Concentration. *Cleanrooms Assoc. Control. Environ.* **2015**, *2*, 37.

(30) Fletcher, N. H. Size Effect in Heterogeneous Nucleation. *J. Chem. Phys.* **1958**, *29*, 572–576.

(31) Na, B.; Webb, R. L. A Fundamental Understanding of Factors Affecting Frost Nucleation. *Int. J. Heat Mass Transfer* **2003**, *46*, 3797–3808.

(32) Turnbull, D.; Vonnegut, B. Nucleation Catalysis. *Ind. Eng. Chem.* **1952**, *44*, 1292–1298.

(33) Hobbs, P. V. *Ice Physics*; Oxford University Press: Oxford, 1974; pp 461–571.

(34) Karydis, V. A.; Kumar, P.; Barahona, D.; Sokolik, I. N.; Nenes, A. On the Effect of Dust Particles on Global Cloud Condensation Nuclei and Cloud Droplet Number. *J. Geophys. Res.: Atmos.* **2011**, *116,* D23204.

(35) Ethington, E. F. *Interfacial Contact Angle Measurements of Water, Mercury, and 20 Organic Liquids on Quartz, Calcite, Biotite, and Ca-Montmorillonite Substrates*; 1990; Vol. 90–409.

(36) Vasileiou, T.; Gerber, J.; Prautzsch, J.; Schutzius, T. M.; Poulikakos, D. Superhydrophobicity Enhancement through Substrate Flexibility. *Proc. Natl. Acad. Sci. U. S. A.* **2016**, *113*, 13307–13312.

(37) Dupont, A.; Paanthoen, P.; Lecordier, J. C.; Gajan, P. Influence of Temperature on the Frequency Response of Fine-Wire Thermocouples over the Range (300K-800K) in Airflows. *J. Phys. E.* **1984**, *17*, 808–812.

(38) Mills, A. F. *Heat Transfer*, Second Edi.; Prentice Hall Englewood Cliffs, NJ, 1999.

**Figures**

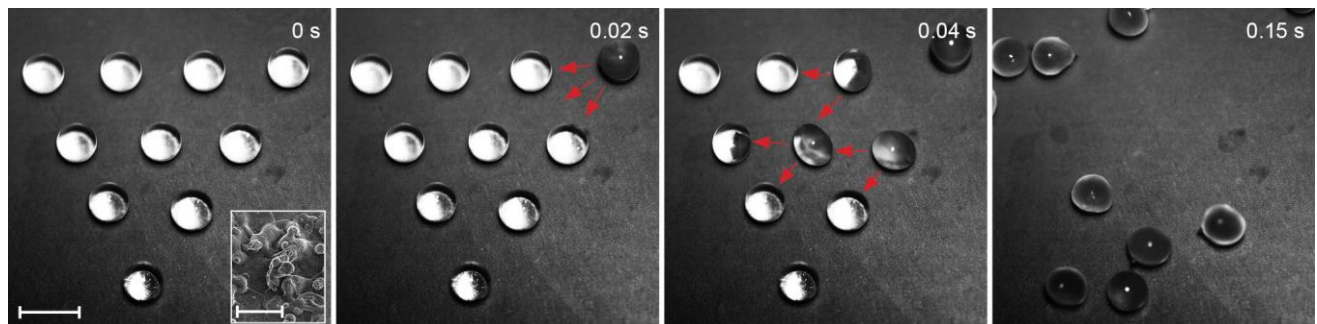

**Figure 1. Top-view image sequence of the rapid, successive freezing of evaporatively supercooled water droplets resting on a micro/nanotextured superhydrophobic surface** (see Video S1). The red arrows illustrate the freezing propagation. The inset shows a micrograph of the surface topography. Scale bar: 4 mm. (inset: 20 μm).

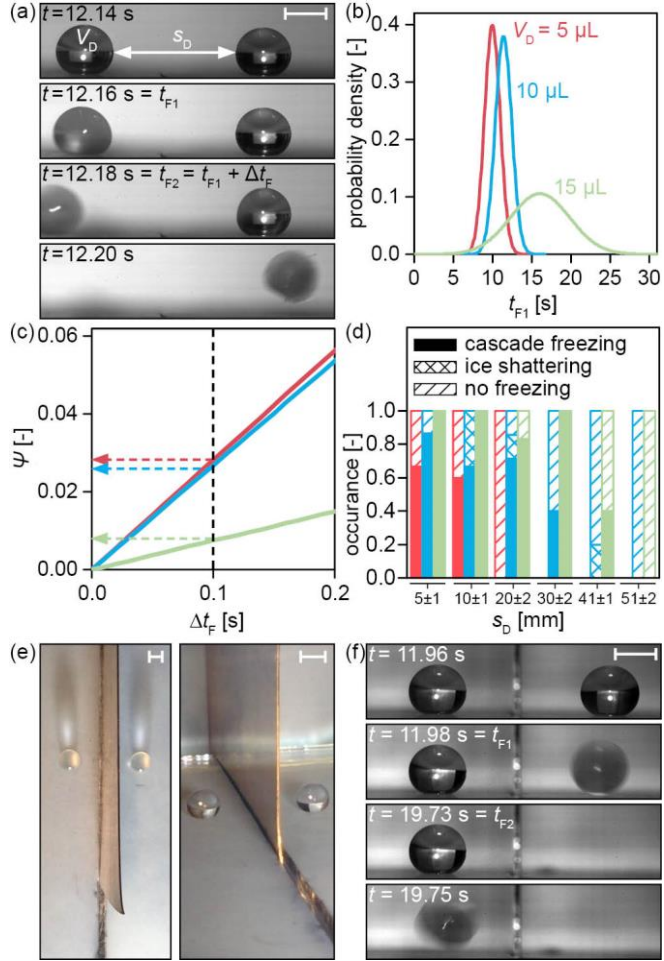

**Figure 2. Freezing interaction of two neighboring droplets.** (a) High-speed image sequence showing cascade freezing of two droplets (initial droplet volume $V_D = 10\,\mu L$) separated by a distance $s_D \approx 5$ mm. Introducing the freezing times $t_{F1}$ and $t_{F2}$, which are the times from reducing the chamber pressure until droplet nucleation of the first (here it is the left droplet) and the second droplet, respectively, and the freezing delay $\Delta t_F = t_{F2} - t_{F1}$ (see Video S3). (b) Probability density function of $t_{F1}$ for three different values of $V_D$, derived from $n = 10$ experiments per $V_D$. (c) Probability $\Psi$ of two droplets independently freezing within a time interval of $\Delta t_F$, computed from data in (b). (d) Experimental outcome for two droplets of the same volume as a function of $s_D$ and $V_D$ ($n = 109$ in total, $n \geq 4$ experiments per combination of $s_D$ and $V_D$): "cascade freezing" is the successive freezing of the two droplets with $\Delta t_F \leq 0.1$ s without any visible ejection of solid ice particles; "ice shattering" is the successive freezing of the two droplets with $\Delta t_F \leq 0.1$ s with visible ejection of solid ice particles due to ice shattering; "no freezing" means that the second droplet did not freeze within $\Delta t_F \leq 0.1$ s. (e) Top- and side-view at an angle of two sessile droplets with a solid metal barrier placed between them. (f) Similar experiment as in (a), but with a solid barrier as shown in (e) placed between the two droplets (see Video S3). Scale bars: (a), (e) and (f): 2 mm. Color code in (b), (c) and (d): red: 5 µL; blue: 10 µL; green: 15 µL.

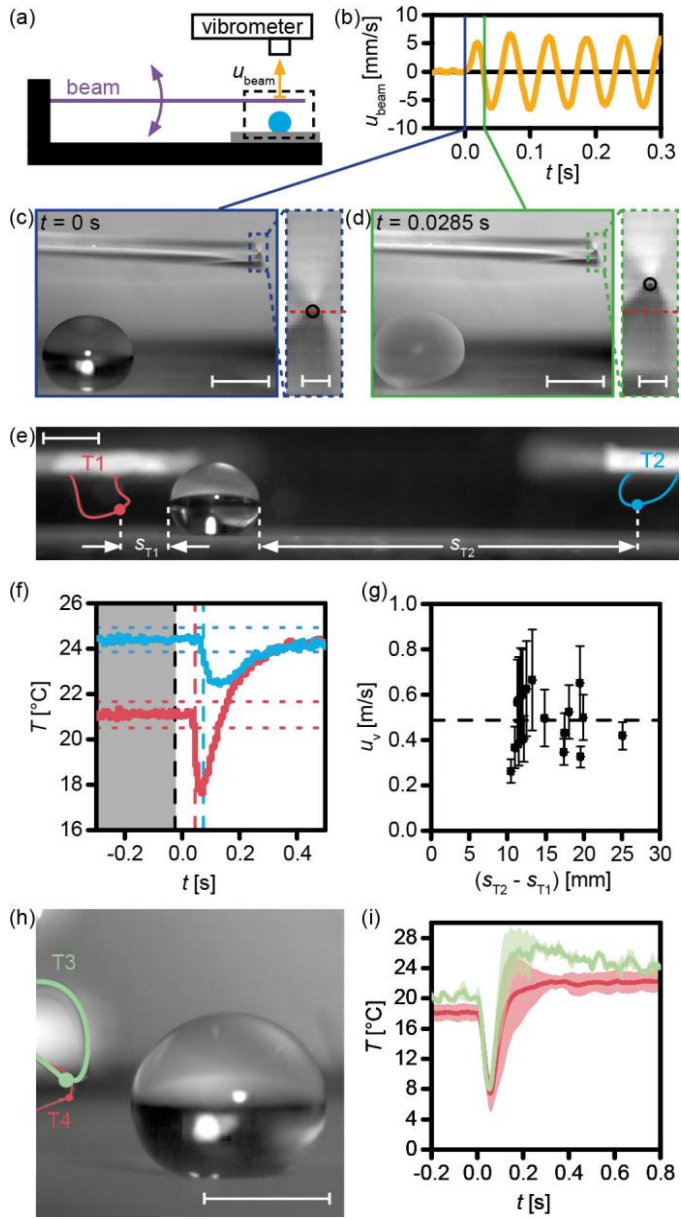

**Figure 3. Vapor release during recalescence freezing.** (a) Schematic of the experimental setup used to characterize vaporization due to recalescence freezing. (b) Beam velocity, $u_{\text{beam}}$, vs time, $t$. (c), (d) Snapshots of the droplet and the tip of the beam (dashed box region in (a)) before and after nucleation, respectively. The dashed red line in the insets shows the undeflected position of the tip. The current tip position is highlighted by a black circle (see Video S6). (e) Experimental setup to measure the velocity of the released vapor, $u_{\text{v}}$, with highlighted thermocouples positioned next to the droplet (T1 and T2). (f) Typical signal recorded by the thermocouples T1 (red) and T2 (blue). The signals in the grey shaded region are used to compute the mean and the standard deviations of the signal. Horizontal dotted lines represent a corridor of plus minus six times the standard deviation around the mean. The vertical dashed lines (red and blue) show the moment when the temperature $T$ leaves the corridor. (g) $u_{\text{v}}$ vs the difference in distance between the droplet and the thermocouples $s_{\text{T2}} - s_{\text{T1}}$ for $n = 18$ experiments. The error bars represent the uncertainty due to the limited recording sampling rate. (h) Setup to measure the vapor temperature $T_{\text{v}}$, using two thermocouples T3 (green) and T4 (red) of different thickness. (i) Averaged ($n = 11$) and response time compensated temperature $T$ obtained by T3 and T4. The minimum of the compensated temperature curve is an estimate for the temperature of the released vapor $T_V \approx 8°C$. The shaded regions represent the error estimate. For all panels, $t = 0$ corresponds to the start of nucleation in the droplet. Scale bars: (c), (d), (e) and (h): 2 mm; Insets in (c), (d): 0.1 mm. Initial droplet volume $V_D = 15\ \mu L$ for all experiments.

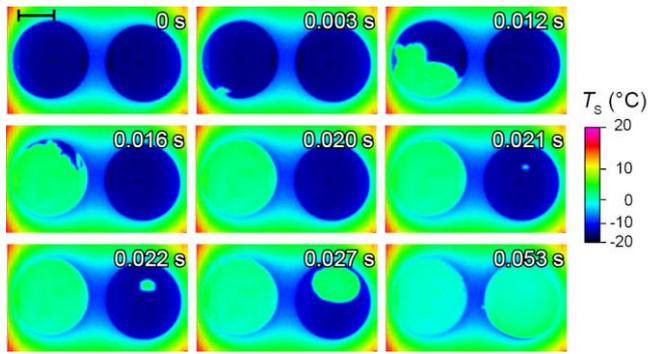

**Figure 4. Freezing propagation mechanism.** Top-view infrared high-speed photography of cascade freezing on PMMA (see Video S7). Initial droplet volume $V_\mathrm{D} = 10\,\mu\mathrm{L}$. Spatial scale bar: 2 mm. Temperature scale bar: Applies only to the droplet, not to the substrate (due to different emissivity).

**Graphical TOC**

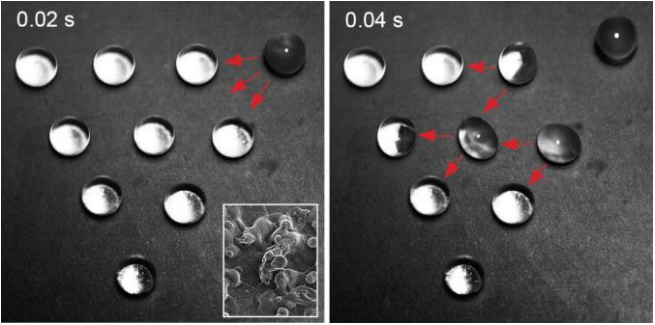